# Tunable band-notched line-defect waveguide in a surface-wave photonic crystal

ZHEN GAO,[1] FEI GAO,[1] YOUMING ZHANG,[1] HONGYI XU,[1] BAILE ZHANG[1,2,*]

[1]Division of Physics and Applied Physics, School of Physical and Mathematical Sciences,

Nanyang Technological University, Singapore 637371, Singapore

[2]Centre for Disruptive Photonic Technologies, Nanyang Technological University, Singapore 637371, Singapore

*Corresponding author: blzhang@ntu.edu.sg

**We propose and experimentally demonstrate a tunable band-notched line-defect waveguide in a surface-wave photonic crystal, which consists of a straight line-defect waveguide and side-coupled defect cavities. A tunable narrow stopband can be observed in the broadband transmission spectra. We also demonstrate that both the filtering levels and filtering frequencies of the band-notched line-defect waveguide can be conveniently tuned through changing the total number and the pillar height of the side-coupled defect cavities. The band-notch function is based on the idea that the propagating surface modes with the resonance frequencies of the side-coupled defect cavities will be tightly localized around the defect sites, being filtered from the waveguide output. Transmission spectra measurements and direct near-field profiles imaging are performed at microwave frequencies to verify our idea and design. These results may enable new band-notched devices design and provide routes for the realization of tunable surface-wave filters on a single metal surface.** © 2016 Optical Society of America

*Surface plasmons.*

http://dx.doi.org/10.1364/OL.99.099999

Surface plasmons (SPs) are electron density waves excited at the interfaces between metal and dielectric materials [1]. Owning to their tightly localized electromagnetic fields, they are deemed promising candidate for the manipulation of photons on subwavelength scales [2-3]. Generally, SPs exist either in the form of surface plasmon polaritons (SPPs) as propagation surface modes [4] or localized surface plasmon polaritons (LSPs) as resonance modes [5]. However, most previous applications of SPPs and LSPs are limited to visible and near-infrared frequency bands, where metals have remarkable inherent loss. To overcome this limitation and export the exciting properties of SPPs and LSPs to low frequencies (middle-infrared to far-infrared, terahertz, and microwave), both the concept of spoof-SPPs [6-9] and spoof-LSPs [10-15] are proposed separately by patterning the extended metal surface or closed metal surface with subwavelength periodic features which can markedly reduce the asymptotic surface plasmon frequencies. More recently, by merging the waveguiding capability of photonic crystals (PCs) [16] and the deep-subwavelength nature of plasmonic metamaterials [6,17], the concept of hybrid locally resonant metamaterials [18-19] or surface-wave photonic crystals [20-24] were developed to manipulate electromagnetic waves in deep-subwavelength scales with superior properties such as broadband high transmission through sharp bends [22], high-Q open resonators [22] and slow wave devices [23].

In this Letter, we go one step further and realize a tunable band-notch function in a surface-wave photonic crystal. We study numerically and experimentally the resonant transmission of surface waves in a line-defect waveguide formed by shortening a row of circular pillars with side-coupled defect cavities in a surface-wave photonic crystal [22]. We observe that a tunable stopband can be generated in the broadband transmission spectra of the line-defect waveguide by properly tuning the cylindrical pillar height and the total number of the side-coupled defect cavities. The transmission spectra are measured and the near-field profiles of the transmission dips are imaged directly at microwave frequencies to verify our design. This band-notched line-defect waveguide has a very simple structure implemented on a single metal surface with tunability in both filtering frequency and filtering level. We believe it can be useful in the future integrated plasmonic circuits as a tunable notch filter in both terahertz and far infrared frequency ranges.

The proposed tunable band-notched line-defect waveguide is constructed on a surface-wave photonic crystal [22], as shown in Fig. 1, which consists of a square array (25 × 25) of circular aluminum pillars with periodicity $d = 5.0$ mm, height $H = 5.0$ mm and radius $r = 1.25$ mm, all standing on a flat metallic surface. Such surface-wave photonic crystals can exhibit complete forbidden band gaps for surface waves [20,24]. Previous Finite Integration Technique (FIT) eigenmode simulation has revealed a complete surface modes band gap from 12.6 GHz to 27 GHz for the current surface-wave photonic crystal [20]. A straight line-defect waveguide can be constructed by shortening the height of a row of aluminum pillars in the middle from $H = 5.0$ mm to $h = 4.3$ mm, as indicated by a long white line in Fig. 1(a). It has been demonstrated that tightly confined surface waves can be guided along the line-defect waveguide in a broad frequency range within the photonic band gap of the surrounding surface-wave photonic crystal [22].

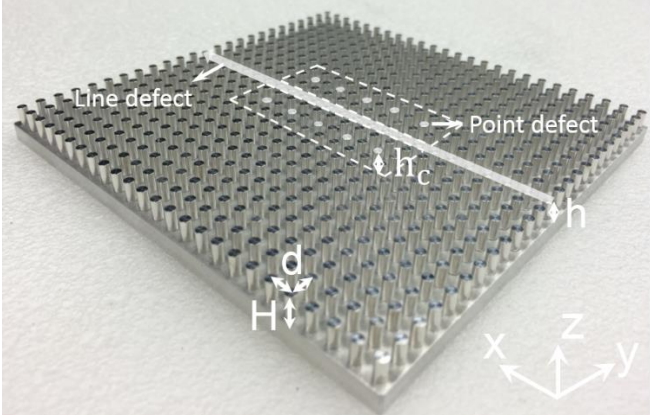

**Fig. 1**. Photography of the tunable band-notched line-defect waveguide in a surface-wave photonic crystal that consists of a straight line-defect waveguide (long white line) with shortened pillar height h and side-coupled point defect cavities (white dots). The surface-wave photonic crystal consists of a square array of circular aluminum pillars with radius $r = 1.25$ mm, height $H = 5$ mm, and periodicity $d = 5$ mm, all standing on a flat aluminum surface. The line-defect waveguide is constructed by shortening a row of pillars from $H = 5$ mm to $h = 4.3$ mm. And the side-coupled point defect cavities (white dots) are created by shorting several side pillars from $H$ to $h_c$.

Now we move on to realize a tunable band notch function in this surface-wave photonic crystal by introducing side-coupled defect cavities with shortened pillar height $h_c = 4.3$ mm along the line-defect waveguide, as shown in Fig. 1. We first simulate the transmission spectra and near-field distributions (E$_z$) of the band-notched filter with different number of side-coupled defect cavities (N = 2, 4, 6, 8, 10, respectively) using the transient solver of CST Microwave Studio with open boundary conditions. A discrete port is placed at the input of the filter as the source. The other discrete port is placed at the output of the filter to detect E$_z$ field as the transmission spectra (S-parameter S21). The simulation results of the transmission spectra detected by the probe discrete port in the frequency range from 12 GHz to 15 GHz are shown in Fig. 2(a). Here, the solid line with different colors are the transmission coefficients of the straight line-defect waveguide without any side-coupled defect cavities (red line), and the band-notched filter with different number of side-coupled defect cavities: N = 2 (green line), N = 4 (blue line), N = 6 (magenta line), N = 8 (cyan line) and N = 10 (olive line), respectively. The introduced side-coupled defect cavities form a very narrow transmission dip in the broadband transmission spectrum of the line-defect waveguide (red line) at almost the same frequency which is exactly the resonant frequency of the side-coupled defect cavities. Note that the small perturbation of the frequencies of the resonant transmission dips is a result of interaction among the point defect cavities and the line-defect waveguide. As the total number of side-coupled defect cavities increase, very narrow transmission dips occur with enhanced filtering levels. It can be seen that the transmission coefficient (S21) is less than -10 dB at the transmission dip of the band notch filter with two side-coupled defect cavities (green line) and further reduces to -28 dB with ten side-coupled defect cavities (olive line). The transmission spectra keep near 0 dB outside the stopband for all band-notched filters. The filtering level of the tunable band-notched line-defect waveguides can be tuned by adjusting the total number of the side-coupled defect cavities. The simulated field distributions of E$_z$ component in a traverse $xy$ plane 1 mm above the top surface of 5-mm-high pillars at the transmission dips of the band-notched line-defect waveguides with different number of side-coupled defect cavities are shown in Fig. 2(b-f), respectively, exhibiting strong field confinement and localization around the side-coupled defect sites. The input surface waves are absorbed and tightly localized by the side-coupled defect cavities one by one and the output surface waves are decreased with increasing the number of side-coupled defect cavities, indicating the tunability of the filtering level of the band-notched line-defect waveguide.

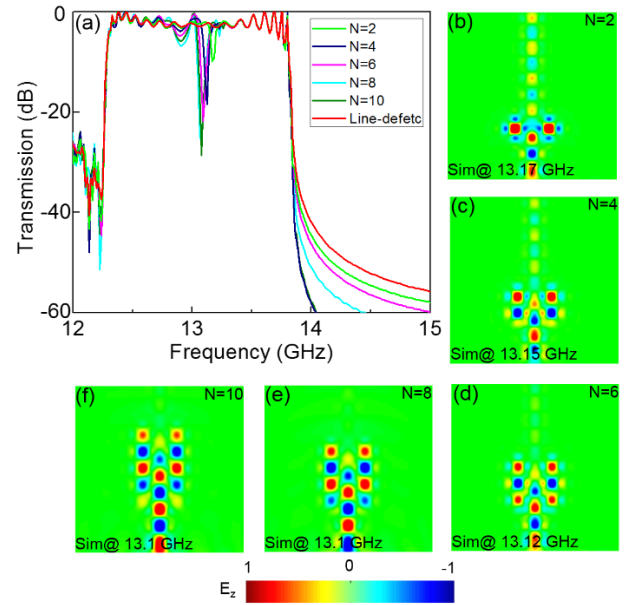

**Fig. 2.** (a) Simulated transmission spectra of the tunable band-notched line-defect waveguides with different number of side-coupled defect cavities (N = 2, 4, 6, 8, 10, respectively). The simulated transmission spectrum of a straight line-defect waveguide without any side-coupled defect cavities (red line) is also shown for comparison. (b-f) Simulated field patterns (E$_z$) of the band-notched line-defect waveguide with different side-coupled defect cavities (N = 2, 4, 6, 8, 10, respectively) in a transverse $xy$ plane 1 mm above the top of 5.0-mm-high pillars of the surface-wave photonic crystal at their resonant transmission dip frequencies.

From now on we fix the number of side-coupled defect cavities as N = 10 and demonstrate the tunability of filtering frequencies of the band-notched line-defect waveguides by tuning the pillar height of the side-coupled defect cavities $h_c$. We use the same simulation method to study three different band-notched line-defect waveguides with three different pillar heights ($h_c$ = 4.2, 4.3, and 4.4 mm, respectively) of side-coupled defect cavities, as schematically shown in the inset of Fig. 3(a). The simulated transmission spectra of the band-notched filter with different pillar heights of side-coupled defect cavities are shown in Fig. 3(a). It can be seen that all three transmission spectra contain one dominant feature: a resonant transmission dip. This feature shifts to lower frequencies as the aluminum pillar height of the ten side-coupled defect cavities increases from $h_c = 4.2$ mm (blue line) to $h_c = 4.3$ mm (olive line) and $h_c = 4.4$ mm (purple line), indicating the tunability of filtering frequencies of the band-notched line-defect waveguides. The simulated transmission spectrum (red line) of a straight line-defect waveguide without any side-coupled defect cavities is also shown in Fig. 3(a) for comparison, which exhibits a broadband transmission from 12.5 GHz to 13.8 GHz without any dips. The simulated $E_z$ field patterns on a transverse *xy* plane 1 mm above the top surface of 5.0-mm-high aluminum rods of the band-notched filter at three different resonant transmission dips are shown in Fig. 3(b-d), respectively. Both the source and probe are indicated with black arrows.

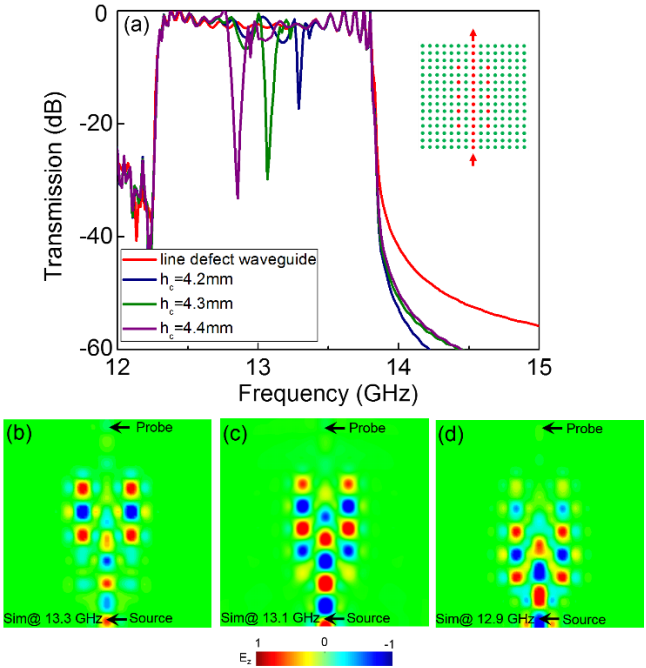

**Fig. 3.** (a) Simulated transmission spectra of the tunable band-notched line-defect waveguides with different pillar heights of side-coupled defect cavities ($h_c = 4.2, 4.3, 4.4$ mm, respectively). The simulated transmission spectrum of a straight line-defect waveguide without side-coupled defect cavities (red line) is also shown for comparison. Inset illustrates the tunable band-notched line-defect waveguide configuration. (b) Simulated field pattern ($E_z$) of a band-notched line-defect waveguide with side-coupled defect cavities pillar height $h_c = 4.2$ mm at the frequency (13.3 GHz) of the transmission dip. (c) Simulated field pattern ($E_z$) of a band-notched line-defect waveguide with side-coupled defect cavities pillar height $h_c = 4.3$ mm at the frequency (13.1 GHz) of the transmission dip. (d) Simulated field pattern ($E_z$) of a band-notched line-defect waveguide with side-coupled defect cavities pillar height $h_c = 4.4$ mm at the frequency (12.9 GHz) of the transmission dip.

We can observe that, at the resonant transmission dip frequencies, the ten side-coupled defect cavities are excited and the propagating electromagnetic energy is gradually absorbed and localized by the side-coupled defect cavities one by one along the propagation direction.

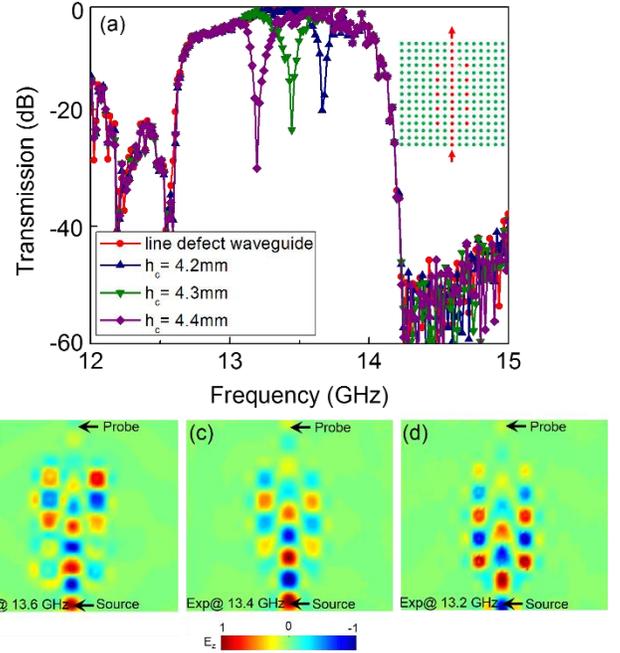

**Fig. 4.** (a) Measured transmission spectra of the tunable band-notched line-defect waveguides with different pillar heights of side-coupled defect cavities ($h_c = 4.2, 4.3, 4.4$ mm, respectively). The measured transmission spectrum of a straight line-defect waveguide without any side-coupled defect cavities (red line) is also shown for comparison. Inset shows the tunable band-notched line-defect waveguide configuration. (b) Measured field profile ($E_z$) of a band-notched line-defect waveguide with side-coupled defect cavities pillar height $h_c = 4.2$ mm at 13.6 GHz. (c) Measured field profile ($E_z$) of a band-notched line-defect waveguide with side-coupled defect cavities height pillar $h_c = 4.3$ mm at 13.4 GHz. (d) Measured field profile ($E_z$) of a band-notched line-defect waveguide with side-coupled defect cavities pillar height $h_c = 4.4$ mm at 13.2 GHz.

To validate our design, we used a vector network analyzer (R&S ZVL-13) to measure the transmission spectra (S-parameters S21) of the band-notched filters in the frequency range from 10 to 15 GHz. Two homemade monopole antennas, serving as the source and the probe, respectively, were placed next to the input and the output of the band-notched line-defect waveguide, as illustrated as red arrows in the inset of Fig. 4(a). We first measured the transmission spectrum (red line in Fig. 4(a)) of a straight line-defect waveguide with shortened pillar height $h = 4.3$ mm. Evidently, the straight line-defect waveguide without any side-coupled defect cavities transmits surface waves over a broad frequency range from 12.8 to 14.1 GHz. Next, we measured the transmission spectra of the band-notched line-defect waveguides with three different side-coupled defect cavity pillar heights. One clear resonant transmission dip can be observed in the transmission spectrum for each band notch filter. The measured near-field distributions ($E_z$) in a transverse *xy* plane 1 mm above the top of 5.0-mm-high aluminum rods of the band-notched filter at three different resonant transmission dips are shown in Fig. 4(b-d), respectively, which match well with the simulation results in Fig. 3(b-d). We can observe that the surface waves are gradually absorbed and tightly localized by the side-coupled defect cavities and filtered from the waveguide output. Note that the simulated transmission spectra (Fig. 3(a)) and the measured transmission spectra (Fig. 4(a)) has a 0.3 GHz shift that is consistent in all transmission bands. This discrepancy mainly comes from the fabrication imperfection. However, as the frequency shift for all transmission bands is about 0.3 GHz, this discrepancy caused by fabrication imperfection is consistent in our measurement.

In conclusion, we have experimentally demonstrated a tunable band-notched line-defect waveguide in a surface-wave photonic crystal at microwave frequencies. Owning to the high tunability of the surface-wave photonic crystal, we were able to achieve different resonant transmission dips with different filtering frequencies and filtering levels in the transmission spectra of the notch filter by tuning the pillar height and total number of the side-coupled defect cavities. Transmission spectra measurements and direct near-field profiles imaging performed at microwave frequencies verify our design. Finally, we would like to emphasize that our design is very general and can be applied to other frequency ranges such as terahertz and infrared frequencies.

**Funding:** This work was sponsored by the NTU-NAP Start-Up Grant, Singapore Ministry of Education under Grant No. MOE2015-T2-1-070 and MOE2011-T3-1-005.